\documentclass[12pt]{iopart}

\usepackage{graphicx}

\begin{document}

\title[Numbers and affinity]{Noise-filtering features of      
    transcription regulation in the yeast {\it S. cerevisiae}}

\author{%
  Erik Aurell$^1$ and
  Aymeric Fouquier d'H\'erou\"el$^{1,2}$ and
  Claes Malmn\"as$^{1}$ and
  Massimo Vergassola$^2$
}

\address{$^1$ Computational Biological Physics, Royal Institute of Technology, AlbaNova University Center, Stockholm, Sweden}
\address{$^2$ CNRS, URA 2171, Institut Pasteur, Dept. ``G\'enomes et G\'enetique'', Research Unit ``G\'enetique in Silico'', 25 rue du Dr Roux, Paris, France}

\eads{%
  \mailto{eaurell@kth.se},
  \mailto{afd@kth.se},
  \mailto{malmnas@kth.se},
  \mailto{massimo@pasteur.fr}
}

\begin{abstract}
  Transcription regulation is largely governed by the profile and the dynamics of transcription factors' binding to DNA. Stochastic effects are intrinsic to this dynamics and the binding to functional sites must be controled with a certain specificity for living organisms to be able to elicit specific cellular responses. Specificity stems here from the interplay between binding affinity and cellular abundancy of transcription factor proteins and the binding of such proteins to DNA is thus controlled by their chemical potential.

  We combine large-scale protein abundance data in the budding yeast with binding affinities for all transcription factors with known DNA binding site sequences to assess the behavior of their chemical potentials. A sizable fraction of transcription factors is apparently bound non-specifically to DNA and the observed abundances are marginally sufficient to ensure high occupations of the functional sites. We argue that a biological cause of this feature is related to its noise-filtering consequences: abundances below physiological levels do not yield signiﬁcant binding of functional targets and mis-expressions of regulated genes are thus tamed.
\end{abstract}


\vspace{1.0cm}
\begin{flushleft}
Running title: \textit{Numbers and affinity}
\end{flushleft}


\maketitle

\section{Introduction}

A major determinant in transcription regulation is the pattern of
transcription factor proteins (TFs) bound in the physical proximity of
the transcribed genomic locus~\cite{Ptashne,Davidson02,Davidson01}.
Intense activity is currently carried out to identify transcriptional
regulatory networks~\cite{PughGilmour2001,Lee,OCT4SOX2NANOG}, their
topology~\cite{Milo2002,Shen-Orr2002,Ping,MBV05} and signs and
strengths of the interactions~\cite{Ronen02}. Specificity is an
obvious need in transcription regulation: functional binding sites
ought to be sufficiently low in energy compared to typical sequences
in the rest of the genome (the so-called background). This energetic
constraint should be coupled with its kinetic counterpart: the TF
should be able to rapidly find its functional targets. Existing
evidence points at a search taking place via 1D sliding along the DNA,
alternated with 3D excursions \cite{Berg3,Marko}. The TF is kept along
the DNA by non-specific electrostatic interactions, recently
characterized experimentally~\cite{Mirny,Gowers}.

Two quantitative variables govern the binding of TFs to DNA: their
cellular abundance and the affinity between the amino acids forming
their binding domains and the various possible stretches of
nucleotides. It has long been recognized in concrete examples that
equilibrium statistical-mechanics models are poised to describe the
binding site occupancy as a function of those parameters, and that
these occupancies are proxies for transcription rates
transcription~\cite{Ptashne,SheaAckers85}.  Detailed models for the
probability of binding to DNA by TFs have recently been reviewed
in~\cite{Bintu2005a,Bintu2005b} and we refer the interested reader
thereto (see also Methods for a concise summary).

\begin{figure}[htp]
  \begin{center}
    \includegraphics[width=16cm]{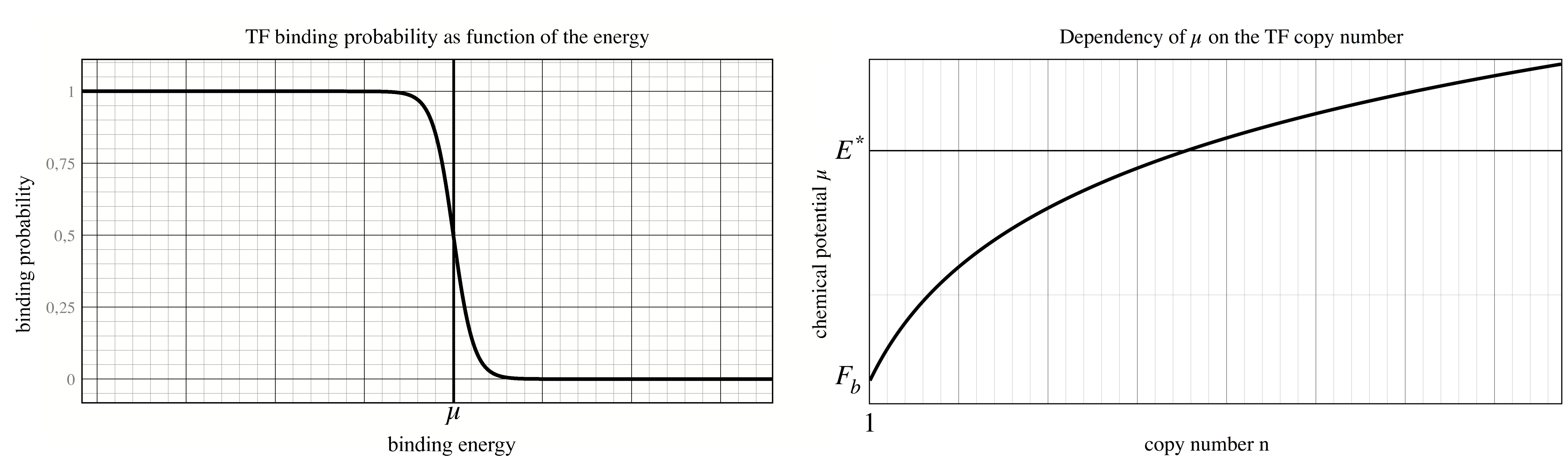}
  \end{center}
  \caption{Left panel: A schematic view of the relation between the
    probability of binding to DNA for a transcription factor and its
    chemical potential $\mu$. Strong binding sites (with energies much
    lower than the chemical potential) have a high occupation
    probability (purple solid line), while the probability to bind
    decreases rapidly as the energy increases. Right panel: the
    relation between the chemical potential and the abundance $n$. The
    background (free) energy $F_b$ is the value of $\mu$ for $n=1$.}
  \label{fig:1}
\end{figure}

The qualitative point of importance here is that the probability of
TF's binding to DNA is controlled by its so-called chemical potential
$\mu$. As illustrated in figure~\ref{fig:1}, strong binding sites (with energy
much lower than $\mu$) are occupied almost certainly, while weak
sites, with energies much higher than $\mu$, are most frequently
empty.  The chemical potential $\mu$ increases with the number of
copies $n$ of the transcription factor as $\log n$ (see,
e.g.,~\cite{Bintu2005a,Bintu2005b}). For a single copy $n=1$, the
value of the chemical potential defines an offset $F_b$, usually
called background energy.  The reason is that $F_b$ controls the
fraction of TF copies bound to DNA either non-specifically or to the
genomic background. Indeed, let $E^*$ denote the minimal binding
energy, i.e. the energy of binding to the consensus sequence of the
TF. From the previous relation $\mu-F_b\propto \log n$, it follows
that if $F_b\simeq E^*$, then the threshold defined by the chemical
potential $\mu$ is larger than (or equals) $E^*$ for any $n\geq 1$. In
other words, even a single copy of the transcription factor would then
be sufficient to ensure persistent binding, at least of the consensus
sequence.  Conversely, as $F_{\rm b}$ becomes less than $E^*$, more
and more TF copies $n$ are needed to have $\mu\geq E^*$,
i.e. persistent occupancy of at least the strongest binding sites. A
minimal abundance (which depends exponentially on the difference
$E^*-F_b$) is then required to have persistent binding of the
strongest sites (supposed to be the functional ones).

\medskip
Detailed quantitative information on the behavior of the chemical
potential for transcription factors of biological interest is
scanty. The relation between binding affinities and abundances was
analyzed in \cite{hwa} for three coliphage TFs (\textit{Mnt},
\textit{CI} and \textit{Cro}) and one bacterial TF
(\textit{LacR}). The result was that the offset $F_b$ is comparable to
the consensus energy $E^*$ for those four TFs. This type of relation
endows the cell with the widest possible window to vary the TF copy
number and differentially regulate various sets of genes. It was
therefore dubbed ``maximum programmability'' \cite{hwa}.

Positing $F_b\simeq E^*$ generally valid seems however too strong a
requirement for the cellular dynamics, as it would make regulation
too prone to errors. In fact, as already
noted in \cite{hwa}, the four TFs which were considered are rather
special: they are all repressors, they operate without much
combinatorial interactions with other factors and their expression is
tightly controlled. This is not the situation encountered in
general. Namely, combinatorial regulation is much more frequent,
especially in eukaryotes, and a large fraction of genes are activated
by TFs to their physiological expression levels. Specificity is not
arising from a single transcription factor but from the sinergistic
and cooperative combination of several factors. We then expect that
the relation $F_b\simeq E^*$, found in \cite{hwa} for four particular
TFs, does not have general validity and that a different relation
holds in the majority of cases. Our goal here is to quantify and
support this expectation by analyzing experimental data for a large
set of transcription factors.

A good model organism to quantitatively investigate the previous issue
is the budding yeast {\it S. cerevisiae}. Concentration data in the
log-growth phase~\cite{tf_amount} and large-scale chromatin
immunoprecipitation binding data, as given by~\cite{Lee,Harbison}, are
both available. The intersection of the two data sets leaves us with a
set of 63 TFs. The difficulty to be overcome is that large-scale
experimental data on binding do not directly provide affinities.  {\it
  A priori}, calorimetric methods \cite{Zhang,Takeda} might be
employed to measure the strength of the interaction of a TF with its
binding sites, but these methods have been hard to scale up, and
values are typically not available for a given TF. One is thus forced
to infer affinity matrices {\it in silico}, from a list of
experimentally detected binding sites.  The procedures and the
limitations of these inferences are recalled in the Methods, together
with the basics of statistical models for TF-DNA interactions. Two
different inference methods were employed: the classical maximum
likelihood argument by Berg and von~Hippel~\cite{BergvonHippel87} and
the QPMEME method, recently introduced in~\cite{Marko03}.  Results for
the relation between affinity and TF abundance, for both ways of
determining the binding energies, are presented hereafter. Biological
consequences, in particular for the control of noise in transcription
regulation, are presented in the Discussion.

\section{Results}

Combining the two experimental data sets on abundance \cite{tf_amount}
and chromatin immunoprecipitation \cite{Lee,Harbison}, a set of 63 TFs was
identified. Affinity matrices for those TFs were then inferred as
detailed in Methods, using both the classical maximum likelihood
procedure \cite{BergvonHippel87} and the QPMEME method \cite{Marko03}.
In both cases, the matrices are {\it a priori} determined only up to a
scale factor. In the first case, following~\cite{BergvonHippel87}, the
factor was set to one in units of $k_{\rm B} T$.  In the QPMEME
method, the scale factor was determined as described in the Methods
via a self-consistency condition, based on the experimental
information on TF abundances.  This condition could be satisfied in 41
out of the 63 cases. In the remaining 22 cases no solution could be
found, for reasons that will be presented in the Discussion.

The matrices derived by the two aforementioned methods agree well in
the majority of the 41 cases where both methods could be employed. A
first measure of the agreement between two energy matrices is whether
they give the best binder at each position, which indeed coincides for
26 TFs out of 41. These 26 instances include cases where one TF admits
more than one consensus sequence, but where both matrices agree on at
least one consensus binder at each position. In 14 cases the sets of
consensus sequences agree completely.  For 15 TFs the sets of best
binders of the two matrices at some position are not overlapping,
\textit{i.e.} in at least one position the sets of best binders
differ.

\begin{figure}[htp]
  \begin{center}
    \includegraphics[scale=0.4]{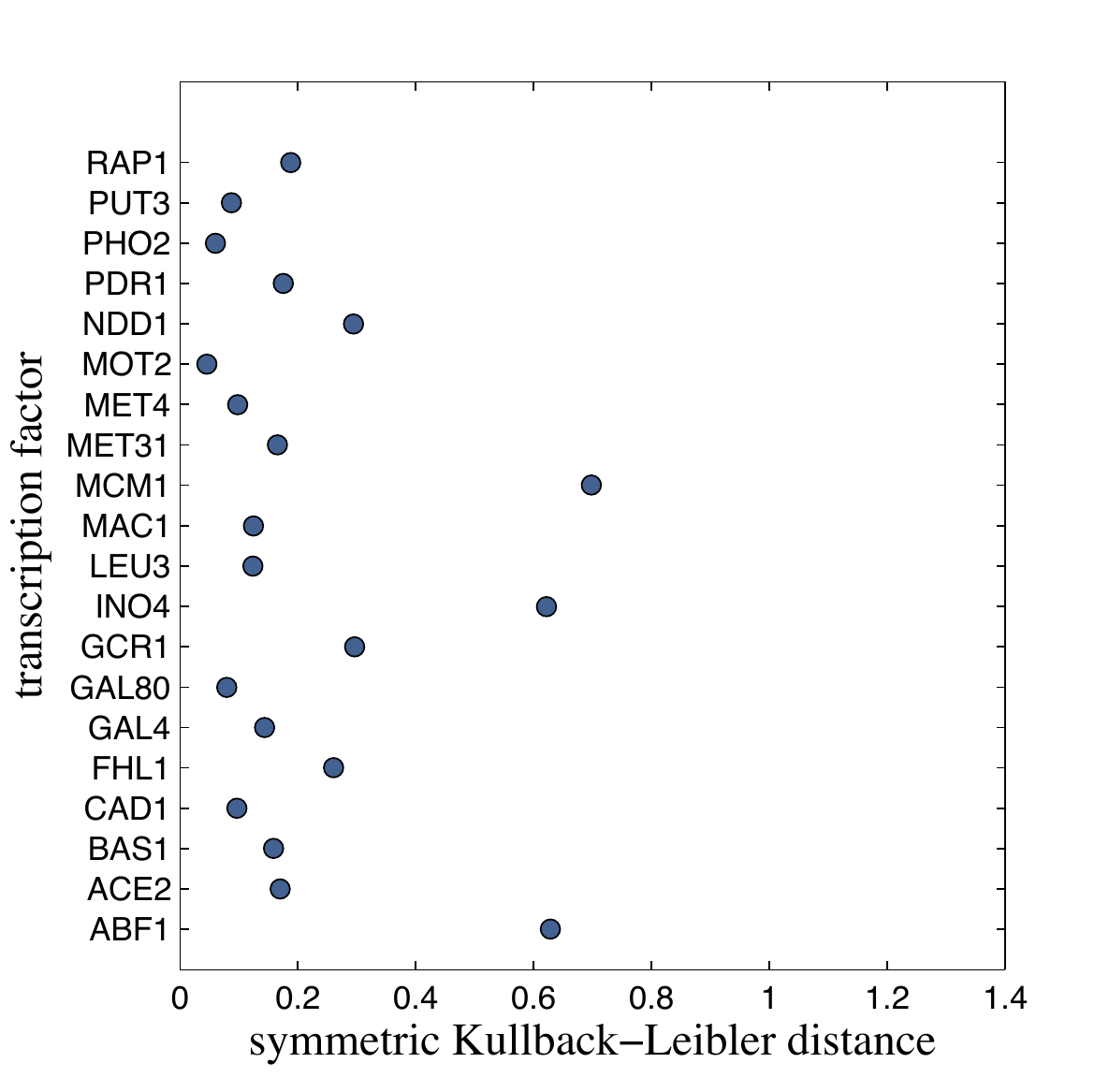}
    \includegraphics[scale=0.4]{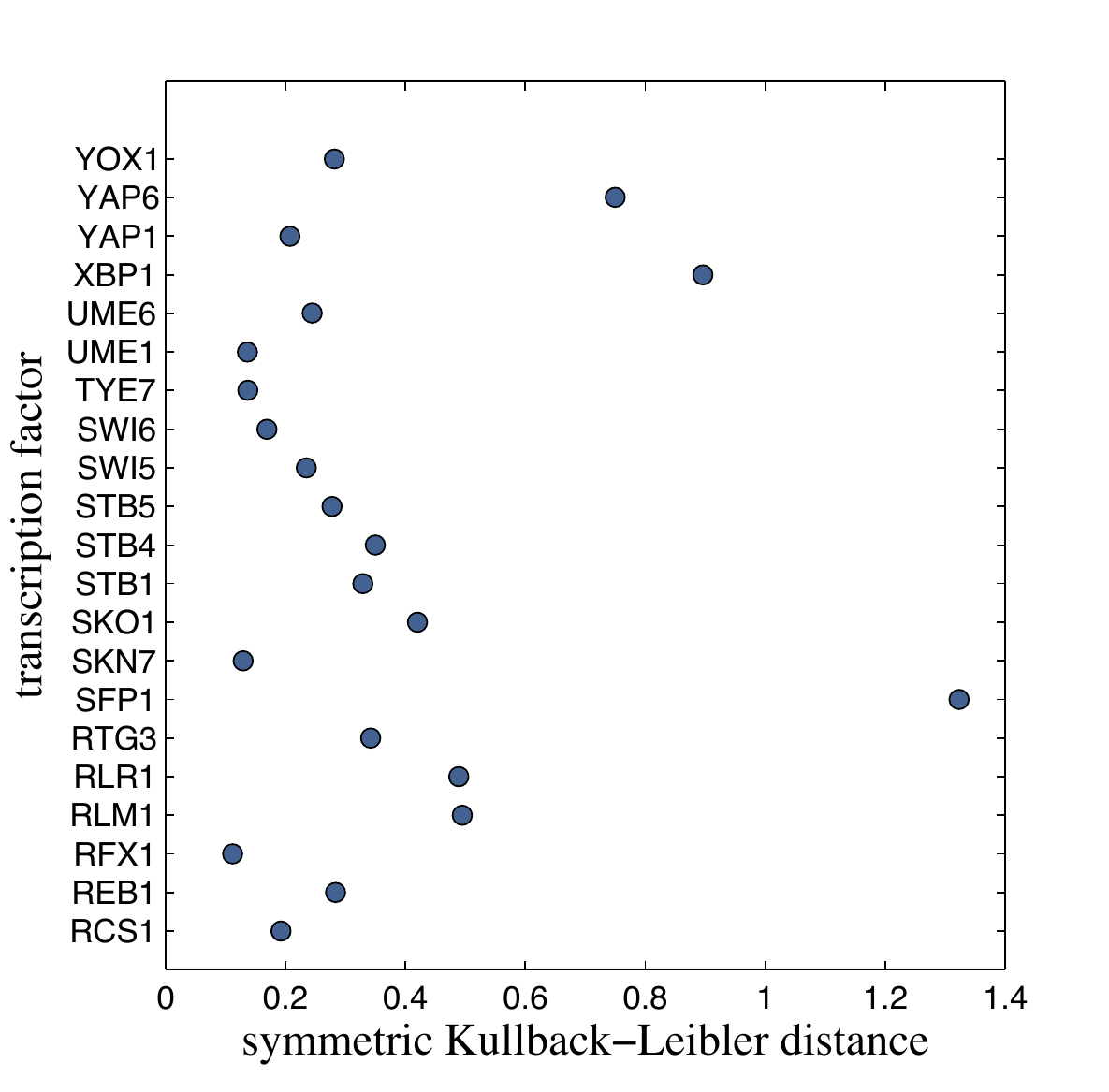}
  \end{center}
  \caption{Average Kullback-Leibler distance per base pair between the
    probability distributions of binding based on computing discrimination
    energies by maximum likelihood arguments~\cite{BergvonHippel87} or
    QPMEME \cite{Marko03} (see also Methods).}
  \label{fig:2}
\end{figure}

A more quantitative comparison, sensitive to the full energy matrix
and not just to the best binder, is to consider the normalized
probabilities $q_{i,\alpha}$, i.e.  the probability that nucleotide
$\alpha$ be found at the position $i$ of the DNA-TF binding
complex. The probabilities computed using the maximum likelihood
procedure \cite{BergvonHippel87} or QPMEME \cite{Marko03} are denoted
by $q^{\rm BvH}_{i,\alpha}$ and $q^{\mathrm QP}_{i,\alpha}$,
respectively. The difference between the two sets of probabilities is
quantified by the symmetric Kullback-Leibler relative entropy
\cite{CT06}\,:
\begin{equation}
S(q^{\mathrm BvH}_i,q^{\mathrm QP}_i) = \frac{1}{2}\sum_{\alpha} 
\left(q^{\mathrm BvH}_{i,\alpha}-q^{\mathrm QP}_{i,\alpha}\right)
\log\frac{q^{\mathrm BvH}_{i,\alpha}}{q^{\mathrm QP}_{i,\alpha}}\,.
\end{equation}
Figure~\ref{fig:2} shows the mean Kullback-Leibler relative entropy per base
pair for the 41 TFs. Except in a few cases, the average differences
per base pair are moderate, on the order of $0.1-0.2$. No correlation
was detectable between the relative entropies and the number of
observed binding sites employed to infer the affinity matrices,
indicating that the differences between the QPMEME and Berg-von~Hippel
matrices are {\it bona fide} fluctuations and not due to finite sample
effects. Detailed properties of the affinity matrices computed using
the two methods are reported in table~1.

\begin{figure}[htp]
  \begin{center}
    \includegraphics[width=8cm]{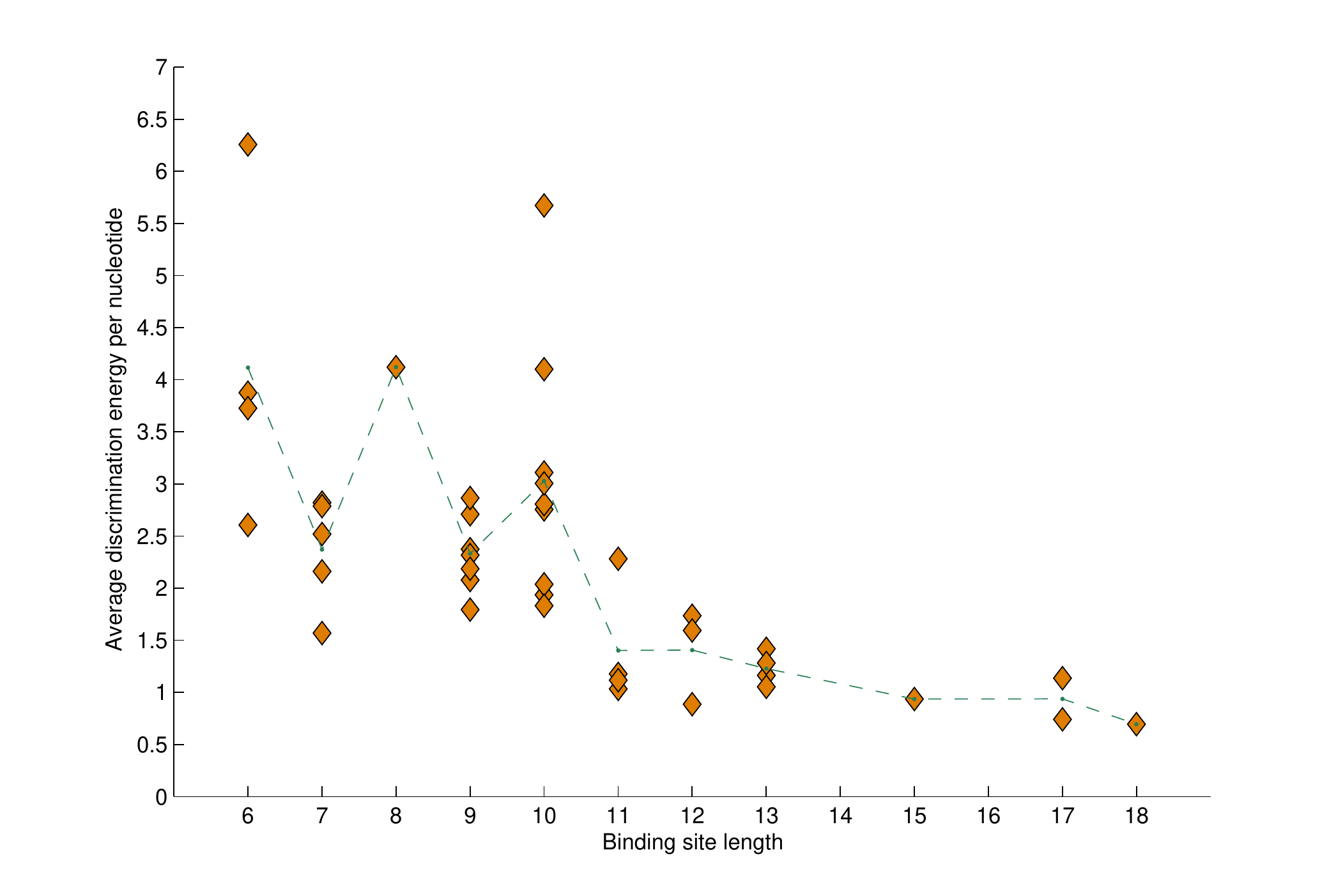}
  \end{center}
  \caption{Average discrimination energy {\it vs} length
    of binding sites.  Reported values refer to energy
    matrices computed using the QPMEME method, as described
    in the Methods.}
  \label{fig:3}
\end{figure}

As a side remark, note that the average discrimination energy per site
generally decreases with the length of the binding site, indicating a
trade-off between these two quantities. Figure~\ref{fig:3} displays the data for
the QPMEME-derived energy matrices; a similar behavior is found for
matrices inferred by maximum likelihood.

\begin{table}[htp]
  \fontsize{8}{8}\selectfont
  \lineup
  \begin{center}
    \begin{tabular}{@{}*{7}{c}}
      \br
      TF name & ${}^{({\bf a})}$ $N_{\rm BS}$ & ${}^{({\bf b})}$ $n_{\rm obs}$ & ${}^{({\bf c})}$ $-r^*$ & ${}^{({\bf d})}$ $|\mu-\langle E_{\rm QP} \rangle|$ & ${}^{({\bf e})}$ $E_{\rm QP}^*-\langle E_{\rm QP} \rangle$ & ${}^{({\bf f})}$ $E_{\rm BvH}^*-\langle E_{\rm BvH} \rangle$ \\ \mr
      ABF1  & 139 & \04818.49 & 1.17 & 12.86 & -15.11 & -30.01 \\ 
      ACE2  & 11 & \0\0538.41 & 1.00 & 17.63 & -17.63 & -12.65 \\ 
      BAS1  & 33 & \0\0861.14 & 1.00 & 23.25 & -23.25 & -15.21 \\ 
      CAD1  & 8 & \0\0622.84 & 1.11 & 17.51 & -19.35 & -14.44 \\ 
      DIG1  & 131 & \01458.26 & 1.44 &  --- & \m--- & -16.37 \\ 
      FHL1  & 75 & \0\0639.38 & 1.22 & 20.60 & -25.08 & -26.41 \\ 
      FKH1  & 89 & \01720.43 & 1.24 &  --- & \m--- & -19.71 \\ 
      FKH2  & 53 & \0\0655.80 & 1.32 &  --- & \m--- & -23.54 \\ 
      GAL4  & 9 & \0\0166.39 & 1.21 & 15.89 & -19.29 & -20.41 \\ 
      GAL80  & 3 & \0\0783.80 & 1.03 & 12.28 & -12.59 & -15.10 \\ 
      GCR1  & 7 & \0\0258.92 & 1.13 & 18.83 & -21.36 & -12.34 \\ 
      GLN3  & 48 & \0\0589.44 & 1.20 &  --- & \m--- & -18.08 \\ 
      HAP5  & 22 & \0\0450.49 & 1.00 &  --- & \m--- & -11.57 \\ 
      INO2  & 27 & \0\0783.80 & 1.17 &  --- & \m--- & -15.26 \\ 
      INO4  & 23 & \0\0521.13 & 1.20 & 34.08 & -41.01 & -18.95 \\ 
      LEU3  & 9 & \0\0124.51 & 1.11 & 18.42 & -20.37 & -15.24 \\ 
      MAC1  & 5 & 14841.76 & 1.03 & 10.71 & -10.98 & \0-8.60 \\ 
      MBP1  & 130 & \0\0521.13 & 1.19 &  --- & \m--- & -20.12 \\ 
      MCM1  & 63 & \08965.92 & 1.24 & 11.06 & -13.67 & -26.07 \\ 
      MET31  & 5 & \0\0521.13 & 1.00 & 16.14 & -16.14 & -10.87 \\ 
      MET4  & 5 & \01295.19 & 1.14 & 12.34 & -14.02 & -15.96 \\ 
      MOT2  & 2 & \04276.97 & 1.03 & 10.31 & -10.63 & \0-8.11 \\ 
      MOT3  & 11 & \01694.68 & 1.00 &  --- & \m--- & \0-9.25 \\ 
      MSN2  & 14 & \0\0124.51 & 1.15 &  --- & \m--- & -14.01 \\ 
      NDD1  & 27 & \0\0799.42 & 1.15 & 16.10 & -18.44 & -21.42 \\ 
      NRG1  & 45 & \0\0555.55 & 1.18 &  --- & \m--- & -17.68 \\ 
      PDR1  & 2 & \01295.19 & 1.03 & 12.51 & -12.93 & \0-7.01 \\ 
      PDR3  & 2 & \0\0166.39 & 1.04 &  --- & \m--- & \0-5.18 \\ 
      PHD1  & 61 & \01417.94 & 1.64 &  --- & \m--- & -15.45 \\ 
      PHO2  & 3 & \06418.52 & 1.09 & 10.40 & -11.37 & \0-8.97 \\ 
      PUT3  & 3 & \0\0736.46 & 1.05 & 11.85 & -12.49 & -15.96 \\
      RAP1  & 70 & \04387.61 & 1.25 & 14.60 & -18.31 & -23.09 \\ 
      RCS1  & 28 & \02733.41 & 1.16 & 21.04 & -24.37 & -16.85 \\ 
      REB1  & 156 & \07514.31 & 1.16 & 13.02 & -15.12 & -23.68 \\ 
      RFX1  & 9 & \0\0376.92 & 1.11 & 14.97 & -16.66 & -18.44 \\ 
      RGT1  & 12 & \0\0194.71 & 1.02 &  --- & \m--- & -12.58 \\ 
      RLM1  & 9 & \0\0736.46 & 1.13 & 24.46 & -27.54 & -14.40 \\ 
      RLR1  & 4 & \0\0521.13 & 1.06 & 19.66 & -20.81 & -11.10 \\ 
      ROX1  & 10 & \0\0238.01 & 1.03 &  --- & \m--- & -13.49 \\ 
      RTG3  & 4 & \01054.72 & 1.00 & 15.63 & -15.63 & \0-7.43 \\ 
      SFP1  & 19 & \0\0258.92 & 1.21 & 46.90 & -56.71 & -17.81 \\ 
      SKN7  & 64 & \02572.11 & 1.25 & 20.63 & -25.77 & -18.55 \\ 
      SKO1  & 8 & \0\0503.71 & 1.00 & 22.35 & -22.35 & \0-9.89 \\ 
      SOK2  & 81 & \0\0314.38 & 1.20 &  --- & \m--- & -16.16 \\ 
      SPT23  & 23 & \0\0432.40 & 1.19 &  --- & \m--- & -13.00 \\ 
      SPT2  & 24 & \01239.69 & 1.25 &  --- & \m--- & -16.26 \\ 
      STB1  & 23 & \0\0319.30 & 1.15 & 27.00 & -31.10 & -17.94 \\ 
      STB4  & 4 & \0\0\098.94 & 1.00 & 18.69 & -18.70 & -10.11 \\ 
      STB5  & 15 & \0\0279.40 & 1.12 & 25.07 & -28.06 & -16.32 \\ 
      STE12  & 147 & \01923.06 & 1.23 &  --- & \m--- & -18.19 \\ 
      SUM1  & 43 & \0\0148.81 & 1.22 &  --- & \m--- & -20.65 \\ 
      SWI4  & 99 & \0\0589.44 & 1.28 &  --- & \m--- & -21.42 \\ 
      SWI5  & 40 & \0\0688.35 & 1.00 & 37.53 & -37.53 & -14.88 \\ 
      SWI6  & 128 & \03335.05 & 1.25 & 26.45 & -32.95 & -19.93 \\ 
      TEC1  & 57 & \0\0529.79 & 1.20 &  --- & \m--- & -15.84 \\ 
      TYE7  & 20 & \0\0486.13 & 1.09 & 19.15 & -20.86 & -18.00 \\ 
      UME1  & 2 & \03037.98 & 1.04 & 11.80 & -12.27 & \0-7.29 \\ 
      UME6  & 63 & \0\0216.63 & 1.21 & 24.83 & -30.04 & -26.54 \\ 
      XBP1  & 2 & \0\0194.71 & 1.00 & 19.74 & -19.74 & \0-5.83 \\ 
      YAP1  & 11 & \01616.86 & 1.06 & 18.47 & -19.66 & -13.73 \\ 
      YAP6  & 3 & \01350.09 & 1.00 & 19.51 & -19.51 & \0-6.87 \\ 
      YAP7  & 48 & \01694.68 & 1.09 &  --- & \m--- & -20.14 \\ 
      YOX1  & 4 & \0\0861.14 & 1.10 & 17.42 & -19.11 & -10.67 \\ \br
    \end{tabular}
  \end{center}
  \caption{Binding parameters for a set of 63 TFs of the yeast {\it S. cerevisiae}, stating numbers of binding sites used in the analysis ({\bf a}), experimentally measured protein abundances ({\bf b}), maximal ratio of binding energy to chemical potential (cf. equation~4 in Methods) ({\bf c}), and in units of $k_{\rm B}T$ the estimates for the chemical potential ({\bf d}) and minimal binding energies (consensus), stemming from both BvH ({\bf e}) and QPMEME matrices ({\bf f}), respectively.}
  \label{tab:tf_info}
\end{table}

Figure~\ref{fig:4} reports the behavior of the background energy $F_b$,
previously defined as the offset of the chemical potential $\mu$ (its
value at unit copy number $n=1$).  The result is that the maximum
programmability relation $F_b\simeq E^*$ proposed in \cite{hwa} is
indeed peculiar to the three coliphage and the bacterial TFs which
were considered. A different behavior is clearly observed in the yeast
{\it S. cerevisiae}. The background energy $F_b$ is not comparable to
the consensus binding energy $E^*$, but is generally smaller and the
difference is correlated with the experimentally observed abundancy
$n_{\rm obs}$, as can be seen in figure~\ref{fig:4}. In other words, the
experimental observations are more in agreement with the behavior $E^*
- F_{\rm b} \propto \log n_{\rm obs}$ than the maximum programmability
relation $E^* - F_{\rm b} \approx 0$.  Note that this holds
irrespective of the method (maximum likelihood or QPMEME) used to
estimate the discrimination matrices.

As a further test, we compared the experimentally measured TF
abundances with the number of binding sites found in
SGD~\cite{SGD-homepage} and as reported by Lee~{\it et
  al.}~\cite{Lee}.  For the latter, we counted all sites of
protein-DNA-interaction with associated $p$-values $<1\cdot10^{-3}$
(L1) and $<5\cdot10^{-3}$ (L5).  The {\it rationale} of this analysis
is as follows.  If the maximum programmability ansatz $F_{\rm b} - E^*
\approx 0$ were satisfied, we should expect that TF abundances are the
main leverage in the control of the number of binding sites. This is
the heuristic advantage provided by maximum programmability \cite{hwa}
and a strong dependence of the number of binding sites on the TF
abundance should then be present.  No such behavior is expected for
the alternative hypothesis $E^* -F_{\rm b}\propto \log n_{\rm obs}$: a
sizable fraction of the TF copies are weakly attached to the DNA, yet
the sites are sufficiently numerous to compete with high-specificity
sites.  A straightforward regression analysis gives coefficients of
regression $R^2$ close to zero, \textit{viz.} $0.0440$ for the SGD set
and $0.0513$ and $0.0900$ for the L1 and L5 sets, respectively. Even
though the p-values for the three sets show some statistical
significance ($0.07,~0.04,~0.006$, respectively), the low values of
$R^2$ indicate that the fraction of the variance explained by the
regression is scanty. To summarize, the correlation between the number
of binding sites and abundance is slightly significant (as should be
expected) but the weakness of the dependency confirms previous
conclusions.

\section{Discussion}
\label{s:discussion}

The integration of binding data provided by chromatin
immunoprecipitation experiments \cite{SGD-homepage,Lee} and abundance
data from~\cite{tf_amount} allowed us to extract information on
the relation between binding affinities and abundances of TFs in
the log-growth phase of the budding yeast {\it S. cerevisiae}.  The
availability of experimental data for other conditions would enable a
wider perspective, yet two main points have already emerged here and
are worth being discussed in their biological consequences and
significance.

\medskip A first technical point is that, while bioinformatic tools to
infer binding free energies generally only give these up to a scale
factor, we have shown that combining the recent method QPMEME
\cite{Marko03} and abundance data can provide an estimate of that
factor. This may be of general methodological interest and useful for
future applications.

For the budding yeast problem considered here, the scale factor could
be estimated for 41 transcription factors out of 63. For the remaining
22 TFs ``individual specificity'' is not ensured by the observed
affinities and abundances, i.e. the binding sites are bound even
though their energy is larger than the chemical potential. This
prevents using QPMEME, since the method works in the strong binding
regime and supposes that all binding sites have energy below the
chemical potential. Biologically, having binding sites occupied
despite their energies being above the chemical potential does not
pose any contradiction, since additional effects such as other factors
and/or regulations of the chromosomal structure might crucially
contribute to specificity.  Indeed, ChIP data (see figure~2 in
\cite{Lee}) clearly indicate that many genes of {\it S. cerevisiae}
are regulated by multiple TFs. Furthermore, global chromatin
remodeling effects will reduce the effective size of the genome which
is accessible to TFs and increase specificity.  Finally, in eukaryotes
it is well known that combinatorial regulation is widespread
\cite{Davidson01} and its mode of action hinges on strong cooperative
effects among the TFs.  The corresponding loci are often structured so
as to require the synergistic action of various TFs and to remain
unbound and inactive if only one of them is present.  Results of our
analysis are in quantitative agreement with this picture.

\medskip The second and main result of our work is that experimentally
observed abundances are marginally sufficient to ensure strong and
persistent binding of {\it S. cerevisiae} TFs to DNA sites. This is
quantified and supported by the results presented in figure~\ref{fig:4}.  More
technically, the background free energy $F_{\rm b}$ was found to be
negative and proportional to $\log n_{\rm obs}$, where $n_{\rm obs}$
is the abundance experimentally measured in
\cite{tf_amount}. Consequently, the chemical potential $\mu$ remains
below the minimal consensus energy $E^*$ if $n\ll n_{\rm obs}$.  This
implies that a sizable part of the TF copies are ``lost in the
background'' and that the \textit{in vivo} observed binding sites are
only occupied with low probability if the abundance is significantly
lower than $n_{\rm obs}$.

What might superficially appear as a waste, ensures in fact an
effective noise-filtering procedure. Fluctuations in the copy number
of proteins are unavoidable in the molecular world and have been
experimentally demonstrated in various cases (see, e.g.,
\cite{Elowitz-Science-2002}).  A few spurious copies of TFs might be
present in the cell due to a variety of mechanisms, going from delayed
degradations, to leaks or lack of tight regulatory controls and
fluctuations in the expression rates.  In an \textit{E. coli} system,
it has recently been shown that extrinsic effects, over and above
cell-cycle dependent changes in gene copy number, acting \textit{e.g.}
through different concentrations of metabolites, ribosomes and
polymerases, may amount to 35\% fluctuations in gene expression
levels, and may persist over a cell cycle~\cite{Elowitz-Science-2005}.
Intrinsic fluctuations, while persisting for shorter times, are also
significant, at the 20\% level~\cite{Elowitz-Science-2005}.  The
relation $E^* - F_{\rm b}\propto \log n_{\rm obs}$ between the
background affinity energy and the abundance of the transcription
factors shown in figure~\ref{fig:4} ensures an effective way to filter out those
fluctuations and control mis-regulations.

\section{Conclusions}
In conclusion, our results point at the importance of quantitative
effects of abundances in the regulatory dynamics of the cell. In
particular, the abundance-affinity relationship $E^*-F_{\rm b}\propto
\log n_{\rm obs}$ demonstrated here is a powerful control lever to
ensure global coherent responses of the cellular regulatory networks
despite the noisy nature of their individual molecular components.

\section{Methods}

Let us consider a TF that diffuses in a cell containing a genomic
sequence of length $L$. The partition function of specific and
non-specific binding to DNA is
\begin{equation}
 Z_b = \sum_{j=1}^{L} e^{-\beta E(S_j)}
+ L e^{-\beta E_{\rm ns}}\,,
\label{eq:Z-def}
\end{equation}
where $\beta$ is the inverse temperature in units of the Boltzmann
constant $k_{\rm B}$ and $S_j$ is the subsequence of length $l$
starting at position $j$ in the genomic sequence.  In~\eref{eq:Z-def}
we have omitted the contribution from the TF freely diffusing in
cytoplasm, assuming that number to be much smaller than the number of
TFs bound.  $E_{\mathrm ns}$ denotes the energy of the state where the
TF is bound non-specifically to the DNA \cite{Berg3,Marko,Mirny}. From
(\ref{eq:Z-def}), it follows the definition of the effective
background (free) energy $F_{\rm b}$ as:
\begin{eqnarray}
\label{eq:binding-probability-2} 
F_{\rm b} = -\beta^{-1}\log Z_b\,.
\end{eqnarray}

A commonly employed expression for the binding energies $E(S)$ is the
additive \textit{energy matrix} form
\cite{Stormo82,Staden84,Stormo00}:
\begin{eqnarray}
\label{eq:additivity}
E(S) = \sum_{i=1}^l \sum_{\alpha=1}^4 \varepsilon_{i,\alpha} S_{i,\alpha}\,.
\end{eqnarray}
Here, the indicator vector $S_{i,\alpha}$ has entries zero or one
depending on which nucleotide $\alpha$ stands at position $i$ in the
sequence $S$, $\varepsilon_{i,\alpha}$ is the free energy contribution
of nucleotide $\alpha$ at $i$ and $\ell$ is the length of the binding
domain. Even though exceptions are known \cite{Bulyk}, the linear form~\eref{eq:additivity} generally gives a good approximation of the
energy profile \cite{StormoFields98}.

Expression~\eref{eq:binding-probability-2} of the background energy
$F_{\rm b}$ may be approximated by an average over a random
ensemble (background).  The approximation is justified in \cite{hwa}
by a mapping to the Random Energy Model~\cite{Derrida}. As for the
choice of the random ensemble, the simplest background model features
independent nucleotides generated with the average genomic frequencies
$p_{\alpha} (\alpha=A,C,G,T)$, yielding:
\begin{eqnarray}
\label{eq:background}\sum_{j} e^{-\beta E(S_j)}\simeq L \langle
e^{-\beta E} \rangle & \equiv & L \prod_{i=1}^{l} \left[\sum_{\alpha}
p_{\alpha} e^{-\beta \varepsilon_{i,\alpha}} \right]\,.
\end{eqnarray}
It follows that 
\begin{eqnarray}
\label{eq:F_b_interm}
F_{\rm b} \simeq -\beta^{-1}\log \left\{L\int dE~\rho(E)\,e^{-\beta E} +
L\,e^{-\beta E_{\rm ns}}\right\}\,,
\end{eqnarray}
where $\rho(E)=\langle\delta\left(E-\sum_{i,\alpha}
\varepsilon_{i,\alpha} S_{i,\alpha}\right)\rangle$ is the density of
states for the random ensemble.  The background density $\rho(E)$ can
be computed by a saddle point expansion, where the first term is
Gaussian \cite{Marko03}.  Figure~\ref{fig:5} compares the empirical energy
density (obtained by the histogram of the energies measured over the
whole genome) with the Gaussian and the first correction. While the
former alone would not be appropriate (the empirical curve is not
symmetric), the correspondence with the latter is quite fair. For a
few TFs the match is less good, mainly because discretization effects
are more pronounced.

For a TF present with $n$ copies in the cell, the probability that a
sequence $S_i$ be bound by the TF takes the Fermi-Dirac form (see,
e.g.,~\cite{hwa,Bintu2005a,Bintu2005b} for more details):
\begin{eqnarray}
\label{eq:binding_n} \mathcal P(S_i) & = & \frac{1}{1 + e^{\beta
(E(S_i) - \mu)}}\,,  
\end{eqnarray}
with the chemical potential $\mu$ implicitly defined by
\begin{eqnarray} 
\label{eq:mu_relation_1} 
n & = & L \int dE~\left[\rho(E) +
\delta(E-E_{\rm ns})\right] \frac{1}{1 + e^{\beta (E-\mu)}}\,.
\end{eqnarray} 
Equation~\eref{eq:mu_relation_1} simply states that the sum over all
the binding sites, weighted by the probability that a TF is bound
there, equals the copy number of the TF in the system.

\subsection{Inference of binding properties}
\label{s:inference}


A list of binding sites for a wide set of TFs of {\it S. cerevisiae}
was downloaded from the SGD database \cite{SGD-homepage}. The binding
sites were extracted from the intergenic regions identified by
chromatin immunoprecipitation experimental data \cite{Lee} as detailed
in \cite{Harbison}. We retained those TFs for which at least two
binding sites and their abundance were available and processed them as
detailed hereafter.

A proxy of the binding properties of the TFs is provided by the
log-odds ratios based on the classical work \cite{BergvonHippel87}:
\begin{eqnarray} 
\label{eq:BvH} 
\Delta\varepsilon_{i,\alpha} = \frac{1}{\lambda}\log\frac{1+n^*_i}
{1+n_{i,\alpha}}\,,
\end{eqnarray} 
where $n_{i,\alpha}$ is the number of observations of nucleotide
$\alpha$ at the $i$-th position in the binding site and $n^*_i$ is the
number of observations of the most frequently observed nucleotide in
that position. $\lambda$ is an unknown scale factor in units of
$k_{\rm B}T$.

The discrimination energy of a sequence $S$ is defined as the
difference between $E(S)$ and the consensus energy and is hence
directly given by $\Delta\varepsilon_{i,\alpha}$ in~\eref{eq:BvH}.
The scale factor $\lambda$ must be determined from at least one
experimentally measured affinity. In the absence of experimental data,
we have set it to unity (in units of $k_{\rm B} T$), which is a fair
average of the values found for a number of prokaryotic examples
in~\cite{BergvonHippel87}, and concords with bioinformatic
practice~\cite{StormoFields98}.

As a second proxy we have used the recently introduced QPMEME
method~\cite{Marko03}. This also does not give access to the binding
energies as such, but to the ratio of binding energies to a chemical
potential, shifted by the mean free energy of binding of the
corresponding TF:
\begin{eqnarray} 
\label{eq:ratio} 
r \equiv \frac{E-\langle E\rangle}{|\mu-\langle E\rangle|}\equiv
\frac{\hat{E}}{|\hat{\mu}|}
=\sum_{i,\alpha}\hat{\varepsilon}_{i,\alpha}S_{i,\alpha}\,,
\end{eqnarray}
where $\langle\bullet\rangle$ denotes the average over the random
background ensemble defined as before.  The calculation of the matrix
$\hat{\varepsilon}_{i,\alpha}$ boils down to a convex optimization
problem, where the width of the background probability distribution is
minimized under the constraints that all sequences in the training set
be bound.  Note that neither the average energy $\langle
E\rangle\equiv \sum_{i,\alpha} p_{i,\alpha}\varepsilon_{i,\alpha}$ nor
the chemical potential $\mu$ are determined by QPMEME. {\em
  Differences} between pairs of energies, \textit{e.g.} discrimination
energies, are determined up to the {\em scale factor} $|\hat\mu|$.

\begin{figure}[htp]
  \begin{center}
    \includegraphics[width=7.6cm]{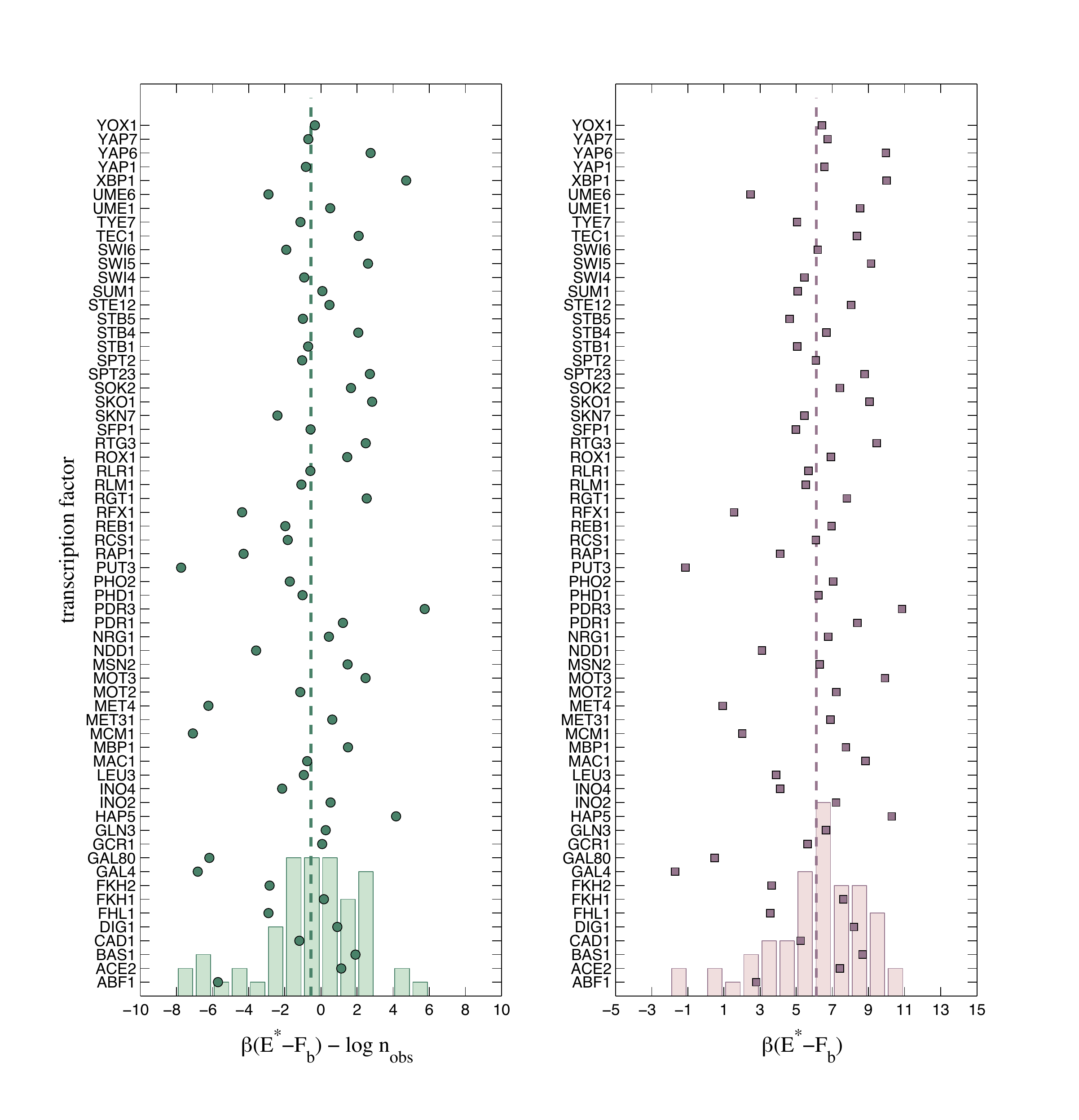}
    \includegraphics[width=7.6cm]{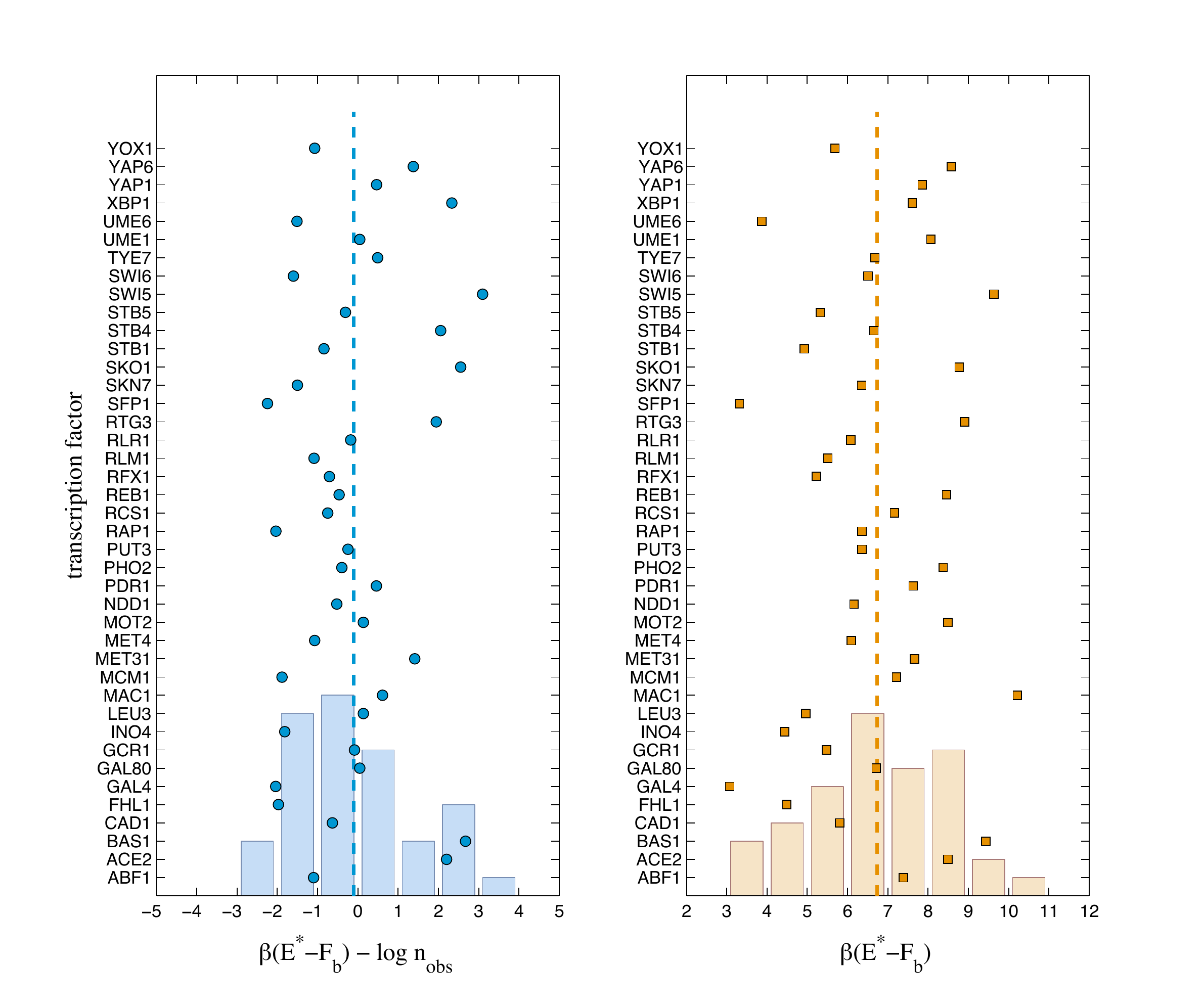}
  \end{center}
  \caption{Comparison of the relation between the background energy $F_b$ and
    the abundance for a set of {\it S. cerevisiae} transcription
    factors. Values of the difference between the consensus energy $E^*$
    and the background energy $F_{\rm b}$ are reported as squares. Their
    values shifted by the logarithm of the TF abundance (as measured
    experimentally) are reported as circles.  Vertical dashed lines
    correspond to the average values for the two sets of points.  Points
    have a sizeable scatter but circles are clearly centered around
    zero.  No relation has been found between the deviation of the
    points around zero and the functional role of the corresponding
    TFs. Long panels: results for log-odds ratio matrices; short panels:
    results for QPMEME matrices.  Histograms give better visual access
    to the distribution widths.}
  \label{fig:4}
\end{figure}

The energy matrices $\Delta\varepsilon_{i,\alpha}$ and
$\hat{\varepsilon}_{i,\alpha}$ have finite sample errors, which could
in principle be estimated as in~\cite{BergvonHippel87}.  Assuming the
sample to be non-biased, these errors decrease with the number of
known binding sites $N_{\rm BS}$ as $1/\sqrt{N_{\rm BS}}$. A
comparison with table~1 reveals that this error is at least on the
order of 10\% (for those TFs for which about a hundred binding sites
are known), ranging up to 50\% (for those with only a few binding
sites known). The chemical potential is determined by the reduced
energy matrix and the observed abundance $n_{\rm obs}$, which also has
experimental errors and is likely to fluctuate {\it in vivo}.  An
estimate of the error in the estimation of the chemical potential is
thus at best on the order of 10\%. This should nevertheless be
sufficient to elucidate statistical trends, which is our purpose here.

The probabilities $q_{i,\alpha}$ appearing in the Results denote the
probabilities that nucleotide $\alpha$ is found at position $i$ in the
TF-DNA complex. They are computed from the energy matrices
$\Delta\varepsilon_{i,\alpha}$ as\,:
\begin{equation}
q_{i,\alpha} = \frac{e^{-\beta\Delta \varepsilon_{i,\alpha}}}
{\sum_{\alpha'} e^{-\beta\Delta \varepsilon_{i,\alpha'}}}\,.
\end{equation}

\subsection{Computing the background free energy}
\label{p:fbcomp}

Definition (\ref{eq:binding-probability-2}) involves two terms: one
describing binding to the genomic background and the other
non-specific electrostatic interactions with the DNA. The latter is
crucial to the target search \cite{Berg3}. As shown in \cite{hwa}, the
background contribution cannot be larger that the non-specific part:
the TF would otherwise diffuse in the background random medium and get
slowed down by its local minima.  In fact, the two contributions are
expected to be comparable.  The division in background and functional
binding sites is indeed dynamical and the former provides the
evolutive reservoir for the latter. Therefore, evolvability of the
regulatory network suggests that the background energy will tend to be
low, compatibly with the aforementioned specificity and kinetic
constraints (see \cite{Lassig1,Lassig2} about evolvability).

\begin{figure}[htp]
  \begin{center}
    \includegraphics{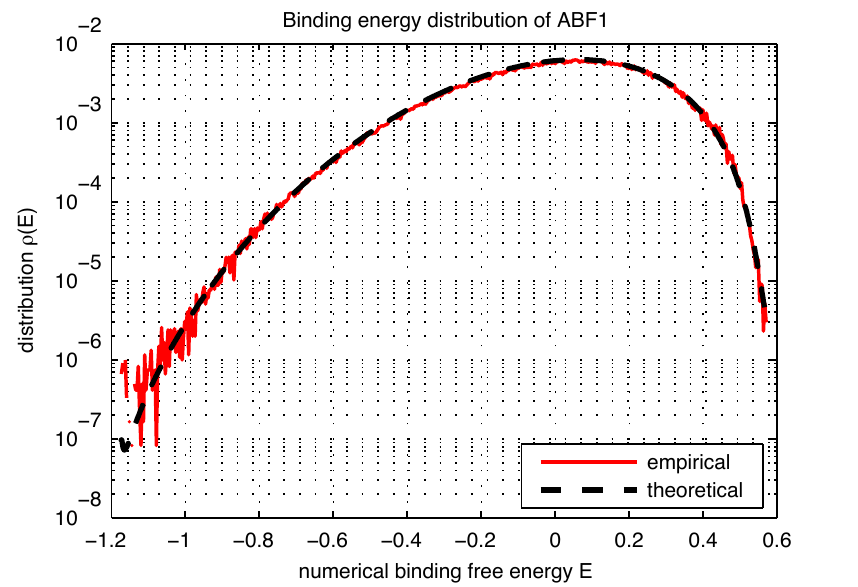}
  \end{center}
  \caption{The density of states for the TF ABF1. Dashed in black,
    the curve obtained for a random background. In red, the empirical
    curve found by computing the distribution of energies over the
    genome. The energy scale has been chosen so as to have the chemical
    potential $\mu=-1$.}
  \label{fig:5}
\end{figure}

Our estimate for the background energy $F_{\rm b}$ in (\ref{eq:F_b_interm})
is then:
\begin{eqnarray} 
\label{eq:F_b_approx} 
\beta\left(E^* - F_{\rm b}\right) = \log\left[ 2L \int dr ~
\rho(r)
e^{-\left(\beta|\hat{\mu}|\right)\left(r-r^*\right)}\right]\,.
\end{eqnarray} Here, $\rho(r)$ is the background
density of states for the energy matrix $\hat{\varepsilon}_{i,\alpha}$
obtained by QPMEME and $r^*$ is the minimal value of the ratio
(\ref{eq:ratio}), that is for the energy $E^*$ of the consensus
sequence(s) $S^*$. The shift to $E^*$ in (\ref{eq:F_b_approx}) is
introduced just to facilitate comparison with the results in figure~\ref{fig:4}.

The quantity $\beta|\hat{\mu}|$ is not determined by the QPMEME method
proper. We estimate it using the relation (\ref{eq:mu_relation_1}),
the fact that in QPMEME binding energies are only determined up to the
relative chemical potential, and the additional information on the TF
abundance $n_{\rm obs}$ from \cite{tf_amount}. Using the previous
arguments on background and non-specific contributions, we get:
\begin{eqnarray} 
n_{\rm obs}=2L \int dr ~
\rho(r) ~ \frac{1}{1 + e^{\beta|\hat{\mu}|
(r+1)}}\,,
\label{eq:mu_relation_2}
\end{eqnarray} 
whence $\beta|\hat{\mu}|$ is extracted and inserted back into
(\ref{eq:F_b_approx}) to obtain the value of the background effective
(free) energy $F_{\rm b}$.  As previously discussed,
(\ref{eq:mu_relation_2}) only has a solution for 41 cases out of 63
TFs. It is instructive to compare with (\ref{eq:mu_relation_1}), which
has a solution for every TF.  The chemical potential $\mu$ then simply
acts as a cut-off, so that sites with energies lower than $\mu$ are
mostly bound, while sites with higher energies are not, and the total
number of bound TFs equals $n_{\rm obs}$.  Depending on $n_{\rm obs}$,
the mostly unbound sites could or could not include \textit{in vivo}
observed binding sites \textit{i.e.}, part of the set of sites from
which the maximum likelihood energy matrices have been constructed.
In (\ref{eq:mu_relation_2}), on the other hand, all the \textit{in
  vivo} binding sites must necessarily have binding energy below the
chemical potential, because these are the constraints under which the
QPMEME reduced energy matrix $\hat\varepsilon$ is determined. Hence,
all sites for which the reduced QPMEME reduced energy is below the
threshold $-1$ will be at least half-filled. Each of these is actually
present in the genome with some probability, which leads to a total
expected number of at least half-filled sites.  Therefore,
(\ref{eq:mu_relation_2}) cannot be solved if $n_{\rm obs}$ is low
enough, because the right-hand side has a lower bound. This happens in
about one third of the cases at hand.

\subsection{Maximal programmability}
\label{s:maximals}
In the simplest scenario where the major contribution in
\eref{eq:mu_relation_1} stems from energies where the Fermi-Dirac
weight can be approximated by the Boltzmann factor, one can invert
\eref{eq:mu_relation_1} to obtain
\begin{eqnarray}
\label{eq:mu_relation_3} \mu \simeq \beta^{-1}\log n + F_{\rm b}\,.
\end{eqnarray}
The occupation probability of a site $t$ reads then $P_t =
\frac{1}{1+{\tilde n}_t/n}$, where the threshold concentration
${\tilde n}_t$ is $e^{\beta(E_t - F_{\rm b})}$.  The minimal copy
number required for strong binding (to the consensus) must then be at
least $e^{\beta(E^* - F_{\rm b})}$.

Maximal programmability \cite{hwa} amounts to positing the lowest
(unity) threshold. The approximate equality $F_{\rm b} \approx E^*$
should then hold. One consequence, which motivates the term, is that
the consensus sequence is then half-bound if there is just a single
copy of the TF present in the cell.  Different regulatory elements can
then have threshold set, or programmed, from one, if their sequences
are the consensus sequence, and upwards, independently of a feedback
induced by the actual TF copy number.

\section*{Acknowledgements}
This work was supported by the Swedish Research Council through contract
2003-4614 (E.A., C.M and A.F.d'H.).

\section*{References}
\bibliographystyle{unsrt}
\bibliography{pb-afmv-references}

\end{document}